# Sub-Doppler spectroscopy of Rydberg atoms via velocity selection memory in a hot vapor cell

Esther Butery[1], Biplab Dutta[1], Stephany Santos[1,2], Sergio Barreiro[2], Weliton Martins[2], Horacio Failache[3], Athanasios Laliotis[1,*]

[1]Laboratoire de Physique des Lasers, Université Sorbonne Paris Nord, F-93430 Villetaneuse, France.
[2]Universidade Federal Rural de Pernambuco, Cabo de Santo Agostinho, Pernambuco 54518-430, Brazil.
[3]Instituto de Física, Facultad de Ingeniería, Universidad de la Republica, J. Herrera y Reissig 565, 1130, Montevideo, Uruguay.
*laliotis@univ-paris13.fr

**We study resonance redistribution mechanisms inside a hot vapor cell. This is achieved by pumping atoms on the first cesium resonance, $6S_{1/2} \rightarrow 6P_{1/2}$, and subsequently probing the velocity distribution of the $6P_{1/2}$ population by a linear absorption experiment on the $6P_{1/2} \rightarrow 16S_{1/2}$ or $6P_{1/2} \rightarrow 15D_{3/2}$ transitions at 514nm and 512nm respectively. We demonstrate that despite the existence of thermalization processes, traces of the initial velocity selection survive in *both* hyperfine levels of the intermediate ($6P_{1/2}$) state. This surprising observation, allows us to perform sub-Doppler resolution vapor cell spectroscopy on Rydberg states using a simple pump-probe setup. At high cesium densities, redistribution mechanisms dominate, causing the velocity selection to vanish. However, analysis of Doppler broadened spectra provides information on the collisional shifts and broadenings of the probed Rydberg states.**

Vapor cell spectroscopy of Rydberg alkali states now finds applications in sensing and quantum technologies. Indeed, Rydberg atoms interact strongly with their environment, making them excellent quantum sensors of microwaves [1] or THz fields [2], [3]. Moreover, Rydberg atoms also interact with their neighbors giving rise to collective effects that could be exploited for realizing single photon sources [4]. The most common way of probing Rydberg states in atomic vapor cells with high frequency resolution is based on EIT spectroscopic schemes, which have proved to be effective in macroscopic [5] as well as thin vapor cells of micrometric thickness [4]. Alternatively, electrical readout after ionization can be used with specially designed cells that integrate metallic electrodes [5].

Rydberg atoms can also be excellent tools for precision measurements of dispersion interactions, such as the Casimir-Polder effect between an atom and a macroscopic surface [6] . For example, precision measurements of the Casimir-Polder interaction have been performed with a sodium beam flying through a metallic cavity [7]. Nevertheless, the interpretation of Rydberg-surface interaction experiments inside vapor cells remains so far challenging [4]. For this reason, our group focuses its efforts on utilizing methods such as selective reflection [8] or thin cell spectroscopy [9] for measuring Rydberg-surface interactions. These experiments could allow demonstration of higher-order terms that become important in the extreme near-field when the size of the atomic wavefunction compares to the atom-surface distance [10]. Additionally, Rydberg-surface interactions are important when hybridizing quantum emitters with photonic platforms with applications in quantum technologies [11].

Atom-surface vapor cell spectroscopy often utilizes a pump-probe scheme to measure Casimir-Polder interactions of excited atomic states [9], [12], [13]. Most commonly, a strong laser drives cesium atoms on the first resonance $6S \rightarrow 6P$ and subsequently a weak non-saturating beam is used to perform spectroscopy on a $6P \rightarrow nD$ or $6P \rightarrow nS$ transition (where n is the principal quantum number of the targeted state). Due to resonance exchange collisions between ground state (6S) and intermediate state (6P) atoms, as well as radiation trapping [14], [15] the intermediate state (both hyperfine levels) population is quasi-thermalized at high vapor pressures. These thermalization mechanisms are very efficient for 6P states because of the large dipole moment matrix element of the relevant coupling ($6S \rightarrow 6P$). For higher excited states, such as the 7P, the velocity selection of the pump laser can be preserved even at high atomic densities [16].

The above experiments [9], [12], [13] also require a narrow atomic reference in the volume, i.e. away from the influence of the atom-surface potential. This is usually achieved by saturated absorption that also provides information on the collisional broadening and shift of the atomic lines. However, the dipole transitions towards Rydberg states are weak and hard to saturate, making the implementation of nonlinear schemes challenging. In this case, EIT spectroscopy is usually employed, which requires an intense coupling laser to reach these highly excited states.

Here, we demonstrate a simple way to perform sub-Doppler Rydberg spectroscopy via the intermediate $Cs(6P_{1/2})$ state in a cesium vapor cell. We show that the initial velocity selection (on the propagation axis of the beams), determined by the frequency of the pump laser (that couples to the $6S \rightarrow 6P$ transition), is preserved within *both* hyperfine levels of the $Cs(6P_{1/2})$ state, for pressures ranging roughly between 0.01-1mTorr, despite the existence of interatomic collisions that tend to thermalize the atomic velocities.

Additionally, we provide spectroscopic information on the collisional shift and broadening of the probed Rydberg states, measured for pressures up to 200mTorr.

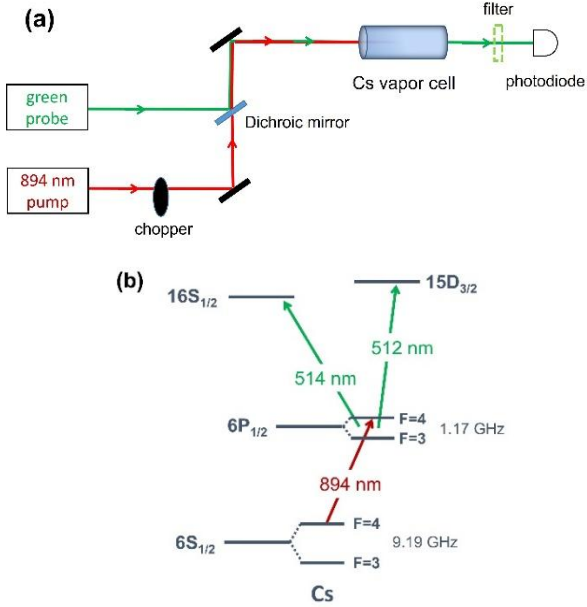

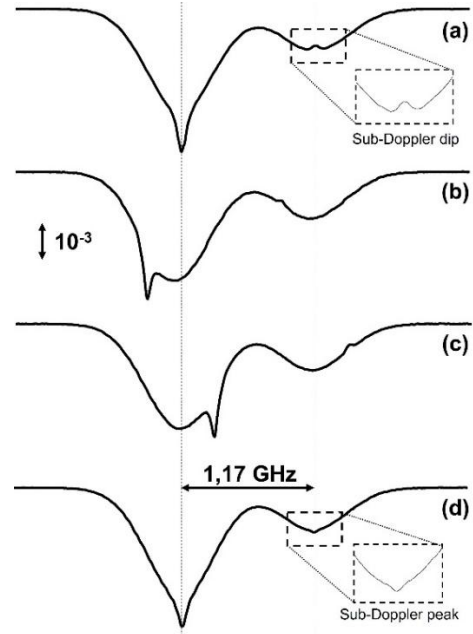

**Fig. 1** (a) Schematic diagram of the experimental set-up used for this experiment. The probe laser can be Frequency Modulated (FM) by applying a voltage on the piezoelectric element of the laser grating and the pump laser is always amplitude modulated (AM) by means of a mechanical chopper. (b) Relevant cesium energy levels for this experiment. The hyperfine structure of the Rydberg levels cannot be resolved in this experiment. The 894 nm laser excites atoms on either the $6P_{1/2}$(F=3) or $6P_{1/2}$(F=4) level.

The basic experimental set-up and protocol of our measurements is described in Fig. 1. The pump source is a diode laser emitting at 894nm, whose beam is amplitude modulated by a mechanical chopper at frequencies ranging between 1-6 kHz. The probe laser is an extended cavity laser whose wavelength can be tuned between 508-514nm (of green color) and can potentially address several S and D Rydbergs states [17]. The two lasers are aligned using a dichroic mirror and they are transmitted through the vapor cell in a copropagating configuration. The vapor cell is 1.0cm long and is made of a metallic body with sapphire windows. It can be heated to temperatures up to 250° C. The pump beam has a radius of about 2 mm and a maximum power of ~5mW. Its frequency is typically stabilized on the atomic resonance with an auxiliary saturated absorption setup (not shown in Fig. 1a). The probe has a radius slightly smaller than the pump (to ensure a good overlap between beams) with a non-saturating power of ~0.5mW. After propagation inside the cell the pump beam is almost totally absorbed. Nevertheless, a filter is used to ensure that no residual pump power makes it to the detector. The probe absorption is typically $10^{-3}$, depending on pump power and vapor pressure. The probe signal is demodulated at the pump AM frequency, isolating the resonant atomic contribution. In parallel with our experiments, we simultaneously record the linear absorption spectrum of the probe laser through a 5cm cell filled with iodine vapor at room temperature. This gives us additional frequency markers, albeit on Doppler broadened transitions, that can be used to calibrate the frequency scale of our scans when necessary.

**Fig. 2** Probe transmission spectra on the $6P_{1/2} \rightarrow 15D_{3/2}$ resonance. For (a), (b), (c) the pump laser is tuned on the $6S_{1/2}$(F=4)$\rightarrow 6P_{1/2}$(F'=4) line with a power of ~1mW (beam radius of 2mm). The pump laser detuning with respect to the resonance is roughly 0, -200 MHz, +200 MHz for (a), (b), (c) respectively. (d) The pump laser is tuned on the $6S_{1/2}$(F=3)$\rightarrow 6P_{1/2}$(F'=4) line with a power of ~1.5mW (beam radius of 2mm). In all cases, the cesium density is on the order of 0.08mTorr.

In Fig.2 we show typical probe transmission spectra for low cesium pressures, about 0.08 mTorr. These results summarize our findings. In Fig.2a the pump is tuned on the $6S_{1/2}$(F=4)$\rightarrow 6P_{1/2}$(F'=4) transition, while the probe scans both $6P_{1/2}$(F'=3,4)$\rightarrow 15D_{3/2}$ transitions. We have experimentally verified that the probe does not saturate the atoms and therefore it simply measures their velocity distribution in both hyperfine levels of the $6P_{1/2}$ state. The population within the $6P_{1/2}$(F'=4) hyperfine level, directly excited by the pump laser, displays a sharp sub-Doppler peak, evidence of the velocity selection imposed by the pump laser, resonant with atoms of $v_z$=0 velocity (where z is the beam propagation direction). The sharp peak sits on a broad pedestal, evidence of excitation redistribution mechanisms such as resonance exchange collisions or radiation trapping [14]. In Fig.2a we can see that the $6P_{1/2}$(F'=3) level, that is not resonant with the IR laser, is also populated due to collisions, whereas radiation trapping should be a less efficient mechanism due to the large frequency split between hyperfine levels. One would expect that the $6P_{1/2}$(F'=3) population would be quasi thermal as reported previously [12]. The surprising observation of the experiments reported here is that a sharp, sub-Doppler dip is observed on the otherwise thermal (Doppler broadened) background, suggesting that the $v_z$=0 population is depleted relative to a Maxwell-Boltzmann distribution. To demonstrate that the observed dip (or peak) in the velocity distribution of the indirectly (or directly) pumped hyperfine level is due to the velocity selection imposed by the pump laser, we perform an additional experiment where the pump frequency is detuned by $\Delta_{IR}$~±200MHz to the blue or red side of the atomic resonance. For these detunings, the pump is resonant with atoms of a velocity $v_z=\pm180$ m/s. The corresponding probe transmission spectra are shown in Fig. 2b and 2c for positive and negative detunings respectively. The apparent shift of the peak in the green probe transmission is $\Delta_{green}=\pm350$ MHz respecting the ratio $\Delta_{green}/\Delta_{IR}=\lambda_{IR}/\lambda_{green}$.

Fig. 2d shows the green laser transmission when the pump laser is tuned on the $6S_{1/2}$ (F=3)→$6P_{1/2}$(F'=4) transition. In this configuration the velocity distribution within the directly pumped hyperfine level displays similar characteristics i.e a sharp sub-Doppler peak due to the velocity selection imposed by the pump sitting on a broad pedestal. However, the population of the $6P_{1/2}$ (F'=3) level now displays a small sub-Doppler *peak* (and not a dip), suggesting an excess of the $v_z=0$ population. As a general rule the amplitude of the Doppler peak (observed in Fig.2d) is smaller than the amplitude of the dip (observed in Fig. 2a). In our experiments the minimum width of the sub-Doppler transition is ~20MHz observed at very low pressures (~0.05mTorr) and low pump power (~0.1mW). This limitation is probably due to probe laser frequency noise at the time scale of the measurement (typically several seconds). For this reason, we cannot resolve the hyperfine split of the probed Rydberg states that is ~1 MHz [18].

To demonstrate the utility of this spectroscopic scheme as a sub-Doppler frequency marker, we perform a frequency modulation (FM) of the probe, allowing us to observe the first derivative of the transmission spectrum. The results are shown in Fig.3 revealing an excellent frequency pointer for the $6P_{1/2}$→$15S_{1/2}$ transition whose width is approximately ~50MHz (limited by the FM modulation). We stress that the above technique can be used as an absolute frequency reference and possibly laser stabilization for all nS and nD atomic states including very highly excited Rydberg transitions.

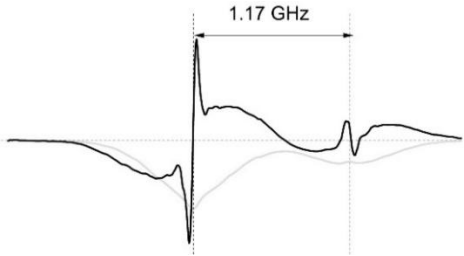

**Fig. 3** FM (black) and direct (gray) transmission spectra on the $6P_{1/2}$→$15D_{3/2}$ resonance. The pump is tuned on the $6S_{1/2}$(F=4)→$6P_{1/2}$(F'=4) line with a pump power of ~1.5mW. The cesium pressure is ~0.15 mTorr. The FM peak-to-peak modulation amplitude (depth) is ~50MHz and the frequency is 233Hz, while the AM frequency is 1.5kHz. FM and AM demodulations are cascaded.

In what follows, we attempt to provide a qualitative explanation of our experimental observations: When tuned on resonance with the $6S_{1/2}$ (F=4)→$6P_{1/2}$(F'=4) transition (Fig.2a), the pump laser burns a hole in the velocity distribution of the $6S_{1/2}$ (F=4) population reducing the number of atoms with zero velocity ($v_z=0$). At the same time, due to optical pumping, atoms with $v_z=0$ are transferred to the $6S_{1/2}$ (F=3) level. When collisional mechanisms start kicking in, for pressures above ~0.01mTorr, population from both hyperfine levels of the ground state, $6S_{1/2}$(F=3) and $6S_{1/2}$(F=4) is transferred to the $6P_{1/2}$(F'=3) level, that is not directly pumped by the laser. However, a transfer from the $6S_{1/2}$(F=4) level should be more efficient as the dipole matrix element of this transition is larger. Therefore, the population of the $6P_{1/2}$(F'=3) presents a dip in its velocity distribution, mirroring the distribution of the $6S_{1/2}$(F=4) level. The inverse scenario is true when the pump is resonant with the $6S_{1/2}$ (F=3)→$6P_{1/2}$(F'=4) transition (Fig.2a). The population of the $6S_{1/2}$(F=4) levels now presents an excess of slow ($v_z=0$) atoms which is subsequently mirrored on the population of the $6P_{1/2}$(F'=3) level that now presents a sub-Doppler peak in the transmission spectrum of Fig.2d. The above also qualitatively explains why the amplitude of the sub-Doppler dip (due to hole burning in the velocity distribution by the laser) is generally larger than the sub-Doppler peak (due exclusively to optical pumping). We stress that the existence of a sub-Doppler peak or dip in the 6P hyperfine level that is not directly pumped by the laser is evidence that its population comes directly from the ground state. This suggests the existence of resonance exchange mechanisms different from direct or velocity changing collisions that have been shown to change the fine structure [19] or even the hyperfine structure [20] of alkali atoms in the presence of a buffer gas. Theoretical studies accounting for both direct and exchange collision kernels [21] can allow a better description [22], [15] of these observations. However, this challenging task, is beyond the scope of this paper.

In Fig. 4 we show a typical probe transmission spectrum at higher atomic densities (10mTorr and 190mTorr). Here, the probe laser is tuned on the $6P_{1/2}$→$16S_{1/2}$ transition and atoms are pumped on $6P_{1/2}$(F'=4). At these high densities any memory of the initial velocity selection, imposed by the pump laser, is erased. Nevertheless, it is possible to study the collisional broadening and shift of the $6P_{1/2}$→$16S_{1/2}$ transition by lineshape analysis. For this purpose, we fit our experimental curves with a Voigt profile, convolving a Gaussian velocity distribution of temperature $T_{col}$ with a Lorentzian of a homogeneous linewidth $\Gamma$. Here, $T_{col}$ represents the effective temperature of the excited state velocity distribution after thermalization due to collisions. Our lineshape analysis shows that $T_{col}\sim450$ K, slightly smaller than temperature of the upper body of the cell that stays fixed at $T_{cell}\sim520$ K throughout the experiment.

Our results for the collisional broadening ($\Gamma$) and shift ($\delta$) are summarized in Fig. 4 (c) for both the $6P_{1/2}$→$16S_{1/2}$ and $6P_{1/2}$→$15D_{3/2}$ transitions. The error bars, ranging between 2-5%, are of statistical nature obtained from linear regression analysis. However, pressure broadening and shift measurements can be sensitive to systematic errors related to the measurement of the cesium reservoir temperature. This could explain the differences from the collisional broadening values reported in [17], performed by selective reflection spectroscopy in a different cesium cell at a lower pressures (5-25mTorr). The collisional broadening and shift reported here, are similar to the values reported in [23] for the equivalent rubidium states. The observed broadening should be predominantly due to collisions of ground state atoms with the Rydberg states as well as collisions between ground state atoms and the intermediate $6P_{1/2}$ state. Similarly, the observed shift is equal to the difference between the shift of the Rydberg state (16S or 15D) and that of the intermediate $6P_{1/2}$ state. Collisional effects on the $6S_{1/2}$→$6P_{1/2}$ transition have been measured via selective reflection [24] giving a broadening of ~1000 MHz/Torr. Although the collisional shift in [24] was negligible, it is suggested that differences could exist between the hyperfine components of the $6P_{1/2}$ state. This could explain the small differences between the observed collisional shift of the two hyperfine components in our experiment (Fig. 3c). In Fig. 4a we notice that the signal amplitude reduces with atomic density. This could be because the effective propagation path of the pump beam is significantly shortened with increasing atomic density. This tends to redistribute the population to many low-lying excited levels due to energy pooling collisions. Collisions with the surface may further quench the excited state population [25].

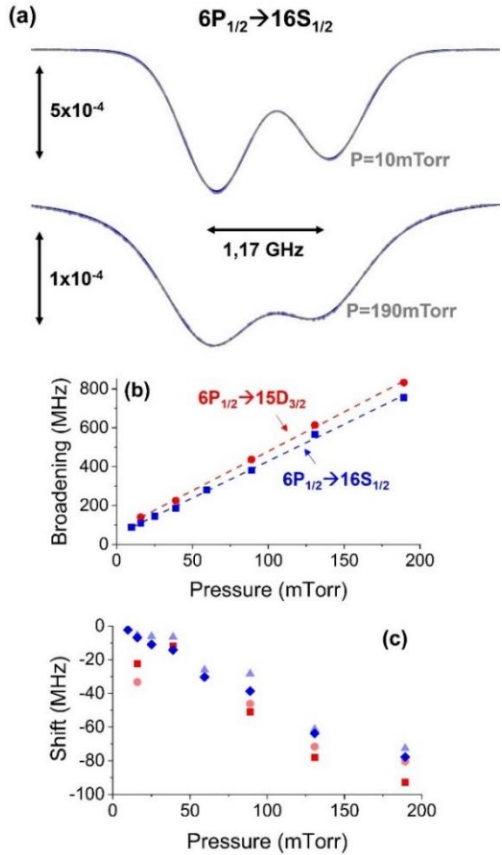

**Fig.4** (a) $6P_{1/2} \rightarrow 16S_{1/2}$ transmission spectrum for 10mTorr and 190mTorr cesium pressures. The experimental spectra (gray lines) are fitted with a Voigt profile (blue lines) to extract the collisional broadening and shift of the transition. (b) Collisional broadening for the $6P_{1/2} \rightarrow 16S_{1/2}$ (blue squares) and $6P_{1/2} \rightarrow 16D_{3/2}$ (red circles) transitions. The slopes of the curves are given by 3900±100MHz/Torr and 4100±100MHz/Torr respectively. The linewidth at zero pressure tends to ~60MHz consistent with experimental observations at low densities. (c) Collisional shifts for the $6P_{1/2}(F'=4) \rightarrow 16S_{1/2}$ (blue rhombi), $6P_{1/2}(F'=3) \rightarrow 16S_{1/2}$ (light blue triangles), $6P_{1/2}(F'=4) \rightarrow 15D_{3/2}$ (red squares) and $6P_{1/2}(F'=3) \rightarrow 16D_{3/2}$ (light red circles) transitions. The slopes are -440±20MHz/Torr, -400±20MHz/Torr for $6P_{1/2}(F'=4,3) \rightarrow 16S_{1/2}$ and -530±40MHz/Torr, -470±40MHz/Torr for the $6P_{1/2}(F'=4,3) \rightarrow 15D_{3/2}$ transition.

In conclusion, we investigate the redistribution of a velocity selected excitation within the first resonant level of cesium ($6P_{1/2}$). We demonstrate that in an intermediate regime of pressures, collisional thermalization is incomplete and therefore, the memory of the velocity selected excitation not only remains within the directly pumped hyperfine level but is also transferred to all hyperfine levels of the $6P_{1/2}$ state. This allows us to perform sub-Doppler pump probe spectroscopy on Rydberg states usually accessible via coherent two-photon processes, such as EIT based schemes, that complicate both the theoretical description and the implementation of the experiment. In principle the linewidth of these narrow resonances is limited by the natural lifetime of the $6P_{1/2}$ state and could be as low as ~10MHz. The signal amplitude is sufficient to allow stabilization of the probe laser (here emitting at ~-508-514nm) on Rydbergs transitions via step-wise spectroscopy. We also obtain information on the collisional broadening and shift of cesium Rydberg states, important for experiments probing atoms close to surfaces [10], [17].

**Disclosures.** The authors declare no conflict of interests.

**Funding.** We acknowledge financial support from the ANR project SQUAT (Grant No. ANR-20-CE92-0006-01), the LABEX FIRST-TF (ANR-10-LABX-48-01) and the LIA Franco-Uruguyen.

**Acknowledgements** The authors are thankful to D. Bloch, I. Maurin, M. Ducloy, J. Rios Leite, M. Tachikawa for discussions.